\documentclass{aa}
\usepackage{graphicx}
\def\ls{\mathrel{\lower0.6ex\hbox{$\buildrel {\textstyle <}
 \over {\scriptstyle \sim}$}}}
\def\gs{\mathrel{\lower0.6ex\hbox{$\buildrel {\textstyle >}
 \over {\scriptstyle \sim}$}}}
\begin{document}
%
   \title{Infrared spectroscopy around 4$\mu$m of Seyfert 2 galaxies: 
    Obscured broad line regions and coronal lines
   \thanks{Based on observations collected at the European Southern Observatory,
   Chile (65.P-0272)}}
   \titlerunning{Infrared spectroscopy around 4$\mu$m of Seyfert 2 galaxies}
   \authorrunning{D. Lutz et al.}


   \author{D.~Lutz
          \inst{1}
          \and
          R.~Maiolino
          \inst{2}
          \and
          A.F.M.~Moorwood
          \inst{3}
          \and
          H.~Netzer
          \inst{4}
          \and
          S.J.~Wagner
          \inst{5},
          E.~Sturm
          \inst{1}
          \and
          R.~Genzel
          \inst{1}
          }

   \offprints{D.~Lutz}

   \institute{Max-Planck-Institut f\"ur extraterrestrische Physik,
              Postfach 1312, 85741 Garching, Germany\\
              email: lutz@mpe.mpg.de, sturm@mpe.mpg.de, genzel@mpe.mpg.de
         \and 
             Osservatorio Astrofisico di Arcetri, Largo Enrico Fermi 5, 
             50125 Firenze, Italy\\
             email: maiolino@arcetri.astro.it
         \and
             European Southern Observatory, Karl-Schwarzschild-Stra\ss\/e 2,
             85748 Garching, Germany\\
             email: amoor@eso.org
         \and
             School of Physics and Astronomy and The Wise Observatory, 
             Tel Aviv University, Tel Aviv 69978, Israel\\
             email: netzer@wise.tau.ac.il
         \and
             Landessternwarte, K\"onigstuhl, 69117 Heidelberg, Germany\\
             email: S.Wagner@lsw.uni-heidelberg.de
             }

   \date{Received 01 July 2002; accepted 20 September 2002}

   \abstract{
      The state of the matter that is obscuring a small circumnuclear region
      in active galactic nuclei can be probed by observations of its broad
      emission lines. Infrared lines are particularly useful since they
      penetrate significant columns of obscuring matter, the properties 
      of which can be constrained
      by comparing infrared and X-ray obscuration. 
      We report on new 4$\mu$m spectroscopy with ISAAC at the ESO VLT of a 
      sample of 12 Seyfert 2 galaxies, probing for broad components
      to the Brackett~$\alpha$ 4.05$\mu$m hydrogen recombination line. Broad
      components are observed in 3 to 4 objects. All objects with a
      broad component exhibit relatively low X-ray obscuring columns,
      and our results are
      consistent with a Galactic ratio of 4$\mu$m obscuration to the BLR and
      X-ray column. In combination with observations of a {\em non}-Galactic
      ratio of {\em visual} obscuration of BLRs and X-ray obscuring column
      in Seyferts,
      and interpreted in a unified AGN scheme, this result can be reconciled
      with two interpretations. 
      Either the properties of dust near the AGN are modified towards
      larger grains, for example
      through coagulation, in a way that significantly flattens the optical/IR
      extinction curve, or the ratio of dust obscuration to X-ray column
      varies for different viewing angles with respect to the axis of
      symmetry of the putative torus.
      Our spectra also provide a survey of emission in the [Si\,IX] 
      3.94$\mu$m coronal line, finding variation by an order of magnitude 
      in its ratio to Br$\alpha$. The first extragalactic detection 
      of the [Ca\,VII] 
      4.09$\mu$m  and [Ca\,V] 4.16$\mu$m coronal lines is reported in the 
      spectrum of the Circinus galaxy.  
            }
\maketitle

\section{Introduction}
\label{sect:intro}
Unified scenarios have been highly successful in explaining several aspects
of the AGN phenomenon, by assuming that different manifestations of the 
AGN phenomenon correspond to similar objects viewed 
from different directions. The detection in polarized light of
broad emission lines in Seyfert 2 galaxies (Antonucci \& Miller 
\cite{antonucci85}, Antonucci \cite{antonucci93}) has been central to the
development of these scenarios. Subsequently, spectropolarimetry has
become the prime tool for detecting hidden broad lines in larger samples
(e.g., Miller \& Goodrich \cite{miller90}; Tran et al. \cite{tran92}; 
Young et al. \cite{young96}; Heisler et al. \cite{heisler97}; Moran et al.
\cite{moran00}; Lumsden et al. \cite{lumsden01}; Tran \cite{tran01}).
X-ray spectroscopy
has been the second key observation, finding Seyfert 2s on average much
more highly obscured than Seyfert 1s and quantitatively establishing the
absorbing column densities in neutral and `warm', i.e. highly ionized
material (e.g., Turner et al. \cite{turner97}; Bassani et al.
\cite{bassani99}). Still, relatively little is known
about the actual distribution and physical state of the material obscuring
the central engine of Seyfert 2 galaxies from our view. Is it in the form
of a compact parsec scale torus (Krolik \& Begelman \cite{krolik86})? Or does 
obscuration on scales of tens or hundreds of parsec play a significant role
(e.g. Maiolino \& Rieke \cite{maiolino95}, Malkan et al. 1998)? 
Mass arguments suggest that at least the Compton-thick absorbers are on
scales of tens of parsecs or less (Risaliti et al. \cite{risaliti99}), but 
lower column components are more difficult to constrain. 
Do outflows contribute to forming the obscuration (e.g. K\"onigl \& Kartje
\cite{koenigl94}; Elvis \cite{elvis00})?
How does the state of the obscuring matter differ from
the interstellar medium in a normal galaxy, given the extreme conditions close
to a powerful AGN? High obscuring columns can be
explained by different scenarios, and the warm dust emission seen in the
mid-infrared cannot uniquely distinguish between compact and more
extended configurations either (e.g., Pier \& Krolik \cite{pier92};
Efstathiou et al. \cite{efstathiou95}; Granato et al. \cite{granato97}).

One way to address some of these issues is to study the obscuration of the
central engine in wavelength ranges that are {\em partially} transparent at the
column densities of interest, and compare the results. Of particular value
are X-rays and the infrared range, but care has to be taken to compare
obscuration towards similar regions. Narrow Line Region (NLR) emission and
mid-infrared dust emission probe regions that are much larger than the 
central X-ray source. Therefore, their obscuration may involve
different foreground material, and cannot be compared directly to
X-ray results. In contrast, reverberation mapping results show the Broad
Line Region (BLR) to be well below a parsec in size
(Netzer \cite{netzer90}), allowing a meaningful
comparison of BLR and X-ray obscuration, by material that might be found in 
both parsec-scale and larger regions. 
Our goal is to constrain infrared obscuration, and thus the state of the
obscuring matter by the detection or nondetection of infrared BLRs in
Seyfert 2s of various X-ray obscuring columns. Recently, Maiolino et al.
(\cite{maiolino01a}) have compared visual and X-ray obscuration in Seyferts, 
finding large deviations from standard Galactic values. In many objects,
the ratio of reddening $E_{\rm B-V}$ and X-ray column $N_{\rm H}$ is low by 
about an order
of magnitude compared to the Galactic value. These results provide additional 
motivation to explore the relation between infrared and X-ray obscuration, 
and compare it with that in the visual.

Practical considerations drive the choice of the
optimal transition for Broad Line Region searches in the infrared. 
Standard interstellar extinction laws have a minimum in the 3-8$\mu$m range, 
then rise 
through the silicate features and drop again steeply beyond 30$\mu$m.
This tends to argue in favour of the  longest wavelength infrared 
observations. However,
the flux of the strongest ($\alpha$) recombination lines drops
approximately with the second power of wavelength, while
dust continuum increases making their detection increasingly difficult.
On the basis of such reasoning and of
ISO spectroscopy of Brackett~$\beta$, Brackett~$\alpha$, and Pfund~$\alpha$
in NGC\,1068, Lutz et al. (\cite{lutz00a}) concluded that Brackett~$\alpha$
with its fairly high line to continuum ratio and low obscuration
is the most promising line for infrared BLR searches. Using
sensitive instruments such as ISAAC on 8m telescopes like the VLT, it is 
now possible to perform such observations and take
a next step beyond earlier infrared searches for BLRs, which either focussed 
on lines that have a good line to continuum but still relatively high
obscuration (Pa$\beta$ 1.28$\mu$m) or are suffering less extinction but are
difficult to measure because of a low line-to-continuum ratio 
(Br$\gamma$ 2.17$\mu$m). Searches
in these lines detected several broad components, but cannot probe beyond
equivalent visual obscurations of 10--20\,mag  
 (e.g. Rix et al. \cite{rix90}; Blanco et al.
\cite{blanco90}; Goodrich et al. \cite{goodrich94};
Ruiz et al. \cite{ruiz94}; Veilleux et al. 
\cite{veilleux97}; Gilli et al. \cite{gilli00}). 

This paper is organized as follows. Sect.~\ref{sect:obs} discusses
the observations and data analysis, Sect.~\ref{sect:coronal}
the result of a coronal line survey obtained from our data, 
Sect.~\ref{sect:bra} presents the results of the Brackett~$\alpha$ 
spectroscopy including line decompositions. We discuss the results and
compare to X-ray and optical work in Sect.~\ref{sect:discussion} and
conclude in Sect.~\ref{sect:conclusions}.      

\section{Observations and data analysis}
\label{sect:obs}
To identify suitable bright Seyfert 2 galaxies we have used
the Bassani et al. (1999) sample with well-known X-ray
obscuring columns. 
We have selected the objects that are brightest in
extinction-corrected [O\,III] $\lambda$5007 
($>$10$^{-12}$erg\,s$^{-1}$cm$^{-2}$), observable from Paranal
(declination $<$20$\deg$), and are at z$<$0.015 to keep Brackett~$\alpha$ in the
observable part of the atmospheric L band. Of the 23 objects in this
parent sample, 12 were observed following right ascension and weather 
constraints during a 
run in March 9-11, 2001. Both the parent sample  and the observed targets
cover a wide range of X-ray obscuring columns. NGC\,1068 was part of the 
parent sample and is included in our discussion, using the existing 
ISO data of Lutz et al. (\cite{lutz00a}; \cite{lutz00b}) observed
with a large $14\arcsec\times\/20\arcsec$ aperture. Due to the prominence of the
central L band peak in NGC\,1068 (Alonso-Herrero et al.~\cite{alonso98}), 
the continuum of these data is still dominated by the central compact peak 
and can be analysed together with the smaller aperture ISAAC sample. 
For part of our sample, spectropolarimetric
observations are available in the literature, directly confirming the 
intrinsic presence of a Broad Line Region. Table~\ref{tab:fluxes} lists
this information, together with some basic source properties and
total line fluxes measured from the spectra.

\begin{figure*}
\center{\resizebox{13.cm}{!}{\includegraphics{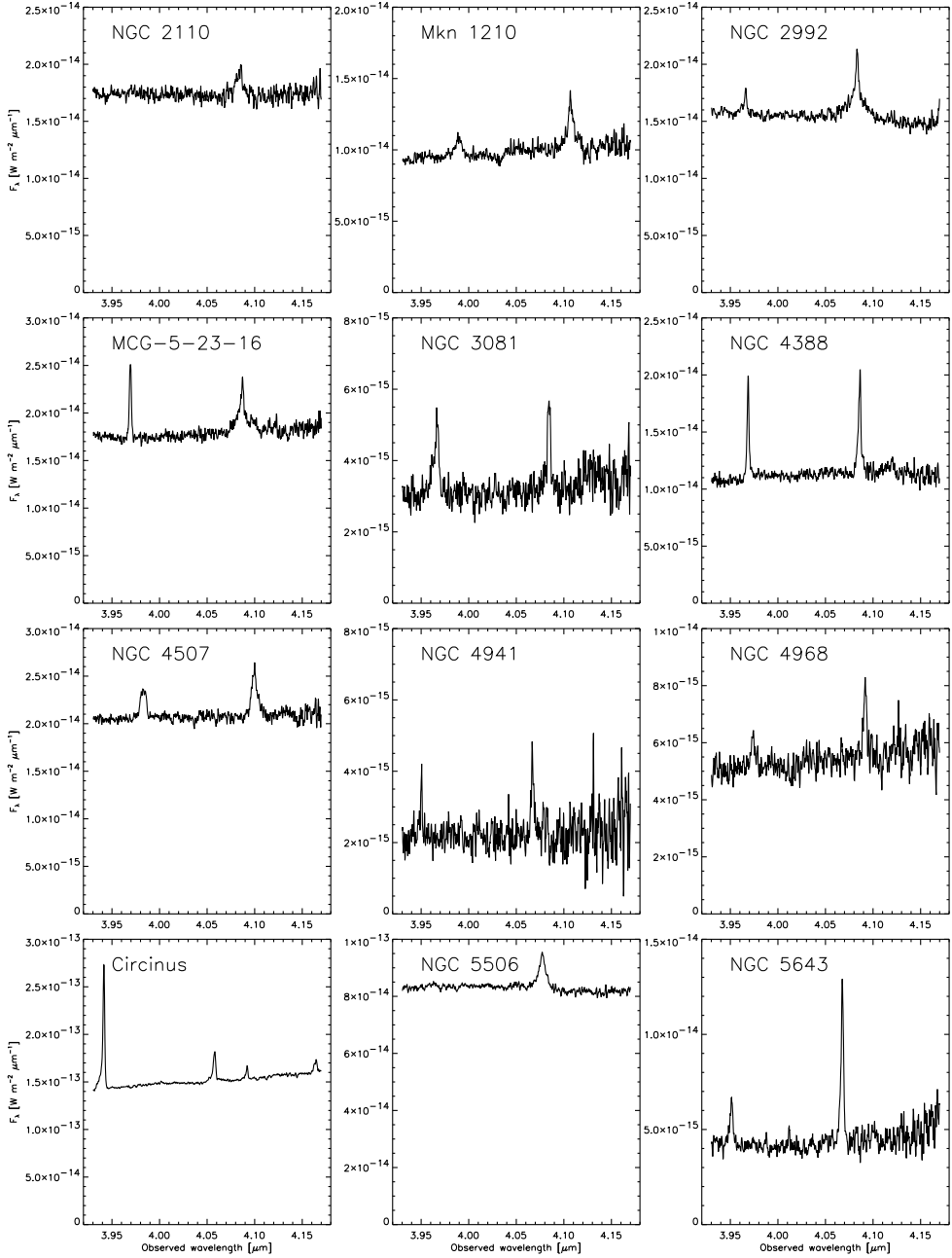}}}
\caption{ISAAC 4$\mu$m spectra of 12 Seyfert 2 galaxies. Noise increases
towards the long wavelength end because of increasing atmospheric opacity.}
\label{fig:blrobs}
\end{figure*}

\begin{figure*}
\center{\resizebox{13.cm}{!}{\includegraphics{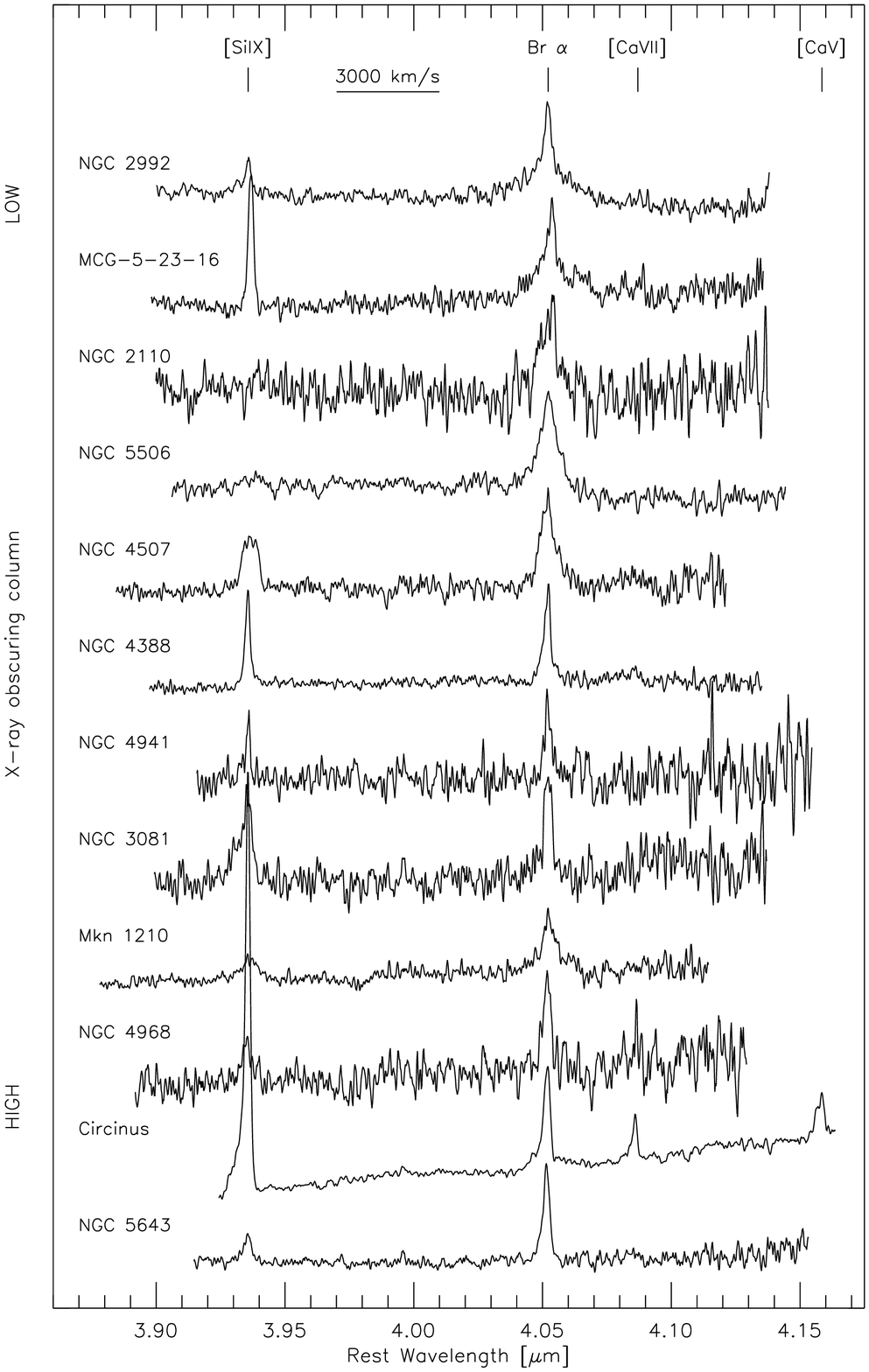}}}
\caption{Spectra sorted in order
of increasing X-ray obscuring column. To facilitate comparison, the
spectra have been shifted to a rest wavelength scale, continuum subtracted
and scaled to the same Br$\alpha$ peak flux.}
\label{fig:blrsort}
\end{figure*}
  
We used the ISAAC-LW medium resolution spectroscopy mode to cover a range 
of 3.93 to 4.17$\mu$m at spectral resolving power $\approx$2500. This range
was chosen because the rapid drop of atmospheric
transmission makes longer wavelengths useless, and to ensure coverage
of the [Si\,IX] 3.94$\mu$m coronal line in all spectra. While we 
present below results for this and other coronal lines obtained from our
spectra, the primary motivation for
including this line was to provide a reference, observed
with the same instrument setting, for the line widths in the Narrow / Coronal
Line Region. These can be used in cases 
where there is ambiguity between a true BLR
and relatively broad NLR components. The bright and compact nuclei were acquired
in the K band and centered in the 1\arcsec\ slit which was oriented
north-south. Observations were done in the chopping and nodding along the slit
scheme suited for thermal infrared observations. Integration times (excluding
overheads) varied
between 37 and 62 minutes per source. Our strategy was to integrate to good 
signal-to-noise in the narrow component of Br$\alpha$. Then, nondetection
of a broad component is meaningful, since in typical Seyfert 1s the broad
line flux is several times brighter than the narrow line flux. In the sample
of Stirpe (\cite{stirpe90}), for example, the median flux ratio of 
broad and narrow H$\beta$ is 20 and the lowest ratio 7, for those 
14 objects that do not exceed a luminosity of $M_{\rm V}=-22$\,mag.

Data reduction followed standard procedures in Eclipse and IRAF. 
The wavelength calibration is based on arc spectra
and on a vacuum scale.
For flux calibration and correction for atmospheric absorptions, we observed
both early type and G type stars. For our particular case of spectra near
Brackett~$\alpha$ with sometimes good continuum S/N, we found early type 
standards little suited because of difficulties to correct for their intrinsic
hydrogen and helium absorption lines, and in some cases emission lines. More 
satisfactory correction of
atmospheric absorption was achieved using early G star spectra which had
first been corrected for their significant intrinsic spectral structure
using the high resolution solar spectrum provided at the ESO ISAAC web pages.
This solar spectrum was convolved to the appropriate resolution and slightly
shifted and scaled, to give optimum cancellation of G star features when 
dividing a
G star and solar spectrum. Since we are interested mostly in the inner
region showing dust and BLR rather than NLR emission we used optimum extraction
of the spectra to get the best signal-to-noise. Effects on the NLR fluxes 
which in principle occur
were verified to be small by inspection of the 2-D spectra and comparison 
of optimum extracted with directly extracted spectra. Only Circinus showed
evidence for faint extended Br$\alpha$ emission peaking $\approx$10-15\arcsec\
from the nucleus, which we did not include in the extracted spectrum.
Figures~\ref{fig:blrobs} and~\ref{fig:blrsort} show the spectra, first as
observed and then sorted by the X-ray obscuring column and scaled in a way 
facilitating comparison. Total fluxes of the coronal lines and of 
Br$\alpha$ (from direct integration of the line profiles) are listed in 
Table~\ref{tab:fluxes}. Due to the narrow slit and optimum extraction scheme,
they correspond to a small $\approx$1\arcsec\ aperture.

\section{A survey of coronal lines}
\label{sect:coronal}

\begin{table*}
\begin{tabular}{lrrrcrrrr}\hline
Source     &cz&[O\,III]&$N_{\rm H}$&BLR in polarized light?&[Si\,IX]&Br$\alpha$&[Ca\,VII]&[Ca\,V]\\
           &km/s&$10^{-14}$Wm$^{-2}$&$10^{20}$\,cm$^{-2}$&                 &
                   \multicolumn{4}{c}{$10^{-20}$\,W\,m$^{-2}$}\\ \hline
NGC\,2992  &2311&0.680&  69&yes (Lumsden et al. 2002)       &  873& 6650&&\\
MCG-5-23-16&2482&0.409& 162&yes (Lumsden et al. 2002)       & 2150& 5010&?&\\
NGC\,2110  &2335&0.321& 289&                                &$<$500&2110&&\\
NGC\,5506  &1853&0.600& 340&? (Tran 2001, Young et al. 1996)& 1000:&11700&&\\
NGC\,4507  &3538&0.158&2920&yes (Moran et al. 2000)         & 2380& 3650&&\\
NGC\,4388  &2524&0.374&4200&yes (Young et al. 1996)         & 2530& 3220&?&\\
NGC\,4941  &1108&0.355&4500&no (Moran et al. 2000)          &  455&  804&&\\
NGC\,3081  &2385&0.215&6600&yes (Moran et al. 2000)         & 1500& 1120&&\\
Mkn\,1210  &4046&0.482&$>10^4$&yes (Tran et al. 1992)       & 1230& 2680&&\\
NGC\,4968  &2957&1.116&$>10^4$&                             &  667& 1060&&\\
Circinus   & 449&6.970&4\,$10^4$&yes (Oliva et al. 1998)    &37700&13900&4040&4610\\
NGC\,5643  &1199&0.694&$>10^5$&no (Moran et al. 2000)       &  870& 2920&&\\
NGC\,1068  &1137&15.86&$>10^5$&yes (Antonucci \& Miller 85) &54000&69000&&\\ \hline
\end{tabular}
\caption{Source properties and integrated line fluxes. 
Heliocentric redshifts are from NED, extinction corrected [O\,III] fluxes
from Bassani et al. (\cite{bassani99}). X-ray obscuring columns
$N_{\rm H}$ are from Bassani et al. (\cite{bassani99}) except for Circinus
(Matt et al. \cite{matt99}). 
}
\label{tab:fluxes}
\end{table*}

\begin{figure}
\resizebox{\columnwidth}{!}{\includegraphics{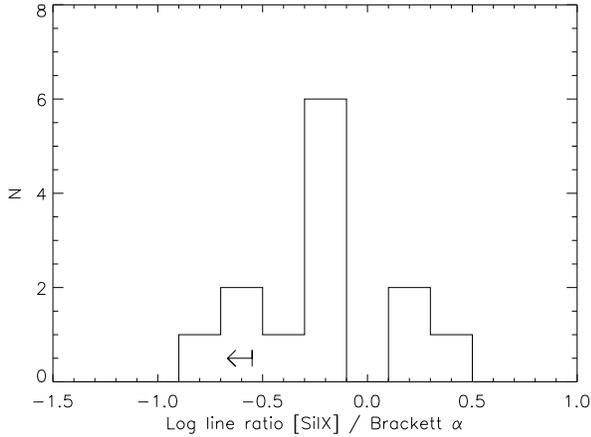}}
\caption{Histogram for the ratio of [Si\,IX]3.94$\mu$m and the narrow component
of Br$\alpha$ in 13 Seyfert galaxies}
\label{fig:sibra}
\end{figure}

The [Si\,IX] 3.94$\mu$m line was first observed in an extragalactic
object by Oliva et al. (\cite{oliva94}) in the spectrum
of the Circinus galaxy. It is one of the highest excitation coronal lines
observed in the spectra of Seyfert galaxies, with a lower ionisation potential
of 303\,eV. It is thus an interesting diagnostic of the coronal line region
of AGN, the density of which can be constrained to be less than about 
10$^6$\,cm$^{-3}$ using the density-sensitive 
ratio [Si\,IX] 2.58/3.94$\mu$m (e.g. Lutz et al.~\cite{lutz00b}).
Our good S/N 4$\mu$m spectra of 12 Seyferts significantly widen the
database of observations of this line. Figure~\ref{fig:sibra} shows
the distribution of the ratio of [Si\,IX] 3.94$\mu$m and the {\em narrow}
(NLR) component of Br$\alpha$, where we have again added NGC\,1068 (Lutz et al. 
\cite{lutz00b}) to our sources. While the [Si\,IX] line is detected in
12 of 13 objects, the observed excitations differ strongly, with the
ratio of [SiIX] and Br$\alpha$ varying by about an order of magnitude, similar
to variations in the strength of optical coronal lines in Seyferts
(e.g., Penston et al. \cite{penston84}, Erkens et al. \cite{erkens97}).

Factors that might bias the observed ratio with respect to the intrinsic NLR
one need consideration. 
Firstly, Br$\alpha$ may be contaminated by starburst 
emission. This effect is likely to be
small in our small aperture. This can be tested on the basis of 
the line profiles. On the high excitation end, for Circinus
both Br$\alpha$ and the coronal lines have similar profiles suggesting a common
origin. Any
starburst contribution to the observed Br$\alpha$ would make the intrinsic NLR 
excitation even higher. On the low excitation end, both [Si\,IX] and Br$\alpha$
in Mkn\,1210 have similar profiles and a high FWHM typical for a NLR, suggesting
that both lines originate in the same region and that the low excitation is 
real. Another low excitation object, NGC\,5643, has similar profiles for both 
lines but
a small linewidth where a starburst contribution is difficult to 
discriminate kinematically. Br$\alpha$ cores that are narrower 
than for the corresponding [Si\,IX] line appear present in
NGC\,3081 and NGC\,4507. These may indicate starburst contamination
but could also partly reflect a NLR where lower and higher excitation species
show different dynamics, as observed in some Seyferts 
(e.g., Appenzeller \& \"Ostreicher \cite{app88}). 
Secondly, underestimating the BLR contribution in a composite 
broad/narrow Br$\alpha$ profile leads
to an underestimate of the [Si\,IX] to narrow Br$\alpha$ ratio. In light of the
Br$\alpha$ profiles and their decomposition discussed in Sect.~\ref{sect:bra},
such decomposition uncertainties might affect NGC\,2110 and NGC\,5506,
which indeed are near the low end of the observed [Si\,IX] to narrow 
Br$\alpha$ range.
Finally, excitation gradients in the NLR may bias small aperture measurements
like ours towards higher excitation. For the extensively observed Circinus
galaxy, comparison to larger
aperture and integral field measurements of coronal lines (Oliva et al.
\cite{oliva94}, Maiolino et al. \cite{maiolino98}) suggests the possible 
effect to be less than a factor 2.

\begin{table*}
\begin{tabular}{lrrrrrrr}\hline
Source     &FWHM (b)& FWHM (n)&$\Delta$v (n)&Flux (b)&Flux (n)&FHWM [Si\,IX]&$\Delta$v [Si\,IX]\\ 
           &km/s&km/s&km/s&\multicolumn{2}{c}{$10^{-20}$\,W\,m$^{-2}$}&km/s&km/s\\ \hline
NGC\,2992  &1850    &  230    & -23&5370    &1300    &240& -14\\
MCG-5-23-16&1450    &  230    & +84&3930    &1110    &190& +77\\ 
NGC\,2110  &        &  618    & -20&$<$3100 &        &&\\ 
NGC\,5506  &1217    &fixed: 460&+39&7150    &5260    &640:&+150:\\ 
NGC\,4507  &        &  540    & -13&$<$4000 &        &520& +74\\ 
NGC\,4388  &        &  260    & -22&$<$3100 &        &190&  -5\\ 
NGC\,4941  &        &  260    &  +0&$<$1600 &        &130& +16\\ 
NGC\,3081  &        &  260    & -16&$<$2300 &        &550& -89\\
Mkn\,1210  &        &  570    &  -1&$<$3900 &        &670& +16\\
NGC\,4968  &        &  270    & -26&$<$2300 &        &400& -31\\
Circinus   &        &  240    & -26&$<$20000&        &170&  +6\\
NGC\,5643  &        &  250    & -35&$<$3100 &        &290&  +4\\ 
NGC\,1068  &        &  640    & -27&$<$80000&        &590&-187\\ \hline
\end{tabular}
\caption{Kinematic data and Br$\alpha$ line decompositions from gaussian 
fits. For objects
without BLR detection, the FWHM of a fit with a single gaussian is given.
Total fluxes from the fits may deviate from the values in Table~\ref{tab:fluxes}
which are based on a direct integration of the profile. $\Delta$v describes 
the shift of the gaussian fitted (narrow) line centroid with respect to the 
galaxy's radial velocity as listed in Table~\ref{tab:fluxes}.}
\label{tab:widths}
\end{table*}

We conclude that these effects do not dominate the [Si\,IX]/Br$\alpha$ spread
of about an order of magnitude among the objects of our sample. 
We also computed simple CLOUDY photoionization models for the `table power 
law' continuum implemented in CLOUDY and varying ionization parameter, showing 
that already modest variations of $\Delta (log\,U) \sim 0.3$ near 
$log\,U = -2$ can change [Si\,IX]/Br$\alpha$ by an order of magnitude.
The observed [Si\,IX] flux is hence a good qualitative
indicator for the presence of an AGN, but a relatively poor quantitative
indicator of the AGN luminosity, due to the considerable variations of observed
excitation.

Two more coronal lines are detected for the first time to our knowledge in 
the spectrum
of an extragalactic source: [CaVII] 4.09$\mu$m and [CaV] 4.16$\mu$m in the 
spectrum
of the Circinus galaxy, at wavelengths in agreement with the ones
inferred by Feuchtgruber et al. (\cite{feuchtgruber01}) from observation
of the planetary nebulae NGC\,6302 and NGC\,7027. Like [Si\,IX] 
and Br$\alpha$, they show
a distinct blue asymmetry, the blue wing extending out to $\approx$800\,km/s. 
A hint of the [Ca\,VII] line is seen in other high excitation spectra 
(NGC 4388, MCG-5-23-16) but at low significance. The rest wavelength of [Ca\,V]
is covered by our spectra for Circinus only.
Detection of these lines (lower ionisation potentials 109 and 67eV) confirms 
again the unique
suitability of the Circinus galaxy for coronal line studies due to 
its brightness, high excitation, and narrow line widths.

Observations of infrared fine-structure and coronal lines can also help 
elucidating the role of dust in forming the line asymmetries and shifts
that are frequently observed in optical coronal lines (e.g., Erkens et al.
\cite{erkens97}). For the prototypical Seyferts NGC\,1068 and NGC\,4151,
these issues are discussed on the basis of ISO data by Lutz et al. 
(\cite{lutz00b}) and Sturm et al. (\cite{sturm99}). They observe some
of the differences between optical and infrared lines that are expected for 
a moderate
amount of dust in the NLR, but also find remaining asymmetries and shifts in
the infrared that must either reflect intrinsic asymmetry of the NLR,
or an optically extremely thick absorber that suppresses even infrared
lines. We list in Table~\ref{tab:widths} the velocity offsets of narrow
Br$\alpha$ and [Si\,IX] with respect to the radial velocity
of Table~\ref{tab:fluxes} (from NED). As expected for our resolving power 
which is 
relatively low for detailed NLR profile studies, both lines are consistent 
with the literature radial velocity in most objects. The significant 
blueshift of high excitation lines in NGC\,1068 has been discussed in detail 
by Lutz et al.
(\cite{lutz00b}). In MCG-5-23-16, both lines are redshifted by $\sim$80\,km/s,
possibly indicating an inaccuracy of the (optical-based) NED redshift.
NGC\,3081 appears to be a case like NGC\,1068, with blueshift and asymmetry
of the coronal line persisting in the infrared. An interesting case deserving
further study is NGC\,4507, where a redshift of the coronal line with 
respect to Br$\alpha$ is suggested.

\section{Results of the Brackett~$\alpha$ spectroscopy}
\label{sect:bra}

In clear cases like NGC\,2992, the broad component of Brackett~$\alpha$
is easily discriminated from the narrow component. At intermediate line widths,
the situation is sometimes ambiguous, however. It is not obvious whether a line
originates in a true BLR of dense clouds close to the AGN, or whether it
represents an unusually wide NLR profile that is seen similarly in the forbidden
and coronal lines (see the example of NGC\,1068; e.g. Lutz et al. 
\cite{lutz00a,lutz00b}). We hence applied the definition of
a broad line as `a component of Brackett~$\alpha$ with FWHM around 
1000\,km/s or more
that is not seen in the forbidden or coronal lines', and used the [Si\,IX]
line as reference. Table~\ref{tab:widths} summarizes our decompositions
from gaussian fits.
In cases where we do not detect a broad component, we quote an approximate
upper limit for its flux determined under the assumption of a representative
value of 3000\,km/s FWHM for the BLR (e.g. Osterbrock \cite{osterbrock77}). 
One should note, however, that the spectral coverage of the
data makes detection of very broad (FWHM 10000\,km/s) lines difficult even 
if the flux were larger.  

A number of objects deserve an individual justification of our decision
to identify a line as originating in a BLR or not, also to indicate 
remaining ambiguities. The Brackett~$\alpha$ line of Mkn\,1210, 
for example, can be plausibly decomposed into a broad and a narrow
component. We decided against a BLR interpretation, however, because the 
[Si\,IX]
line shows wide wings as well, the two profiles being indistinguishable
within the S/N limitations. This argues against the interpretation
of Veilleux et al. (\cite{veilleux97}) who considered Mkn 1210 a BLR
detection (but see caveats in their appendix). Their [Fe\,II] and 
Paschen $\beta$ profiles are very similar
to our [Si\,IX] and Brackett~$\alpha$ profiles, while we do not find
evidence for the FWHM 3000\,km/s component possibly seen in their 
Brackett~$\gamma$ data.

\begin{figure*}
\center{\resizebox{12.cm}{!}{\includegraphics{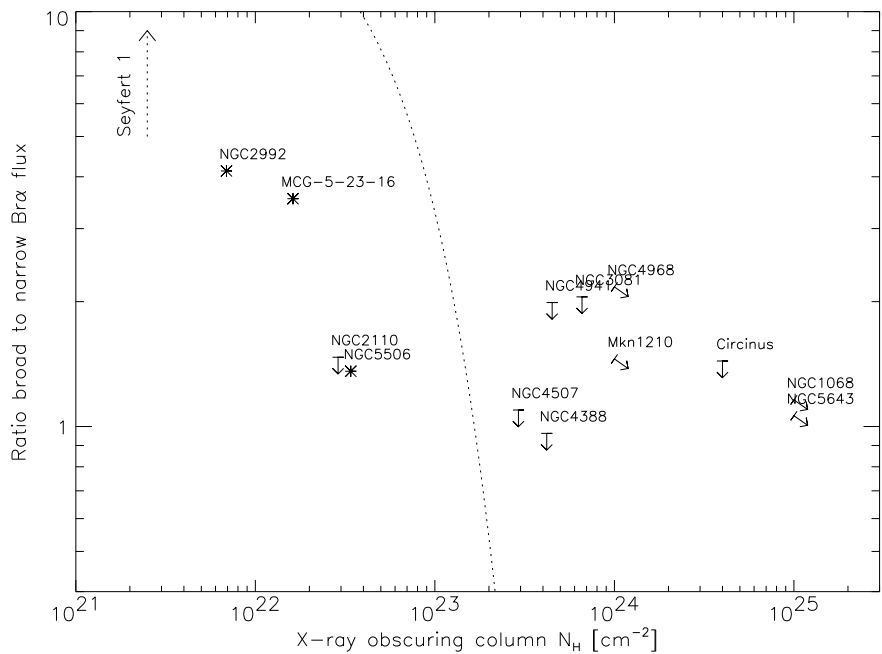}}}
\caption{Flux ratio of broad and narrow components of Brackett~$\alpha$,
as a function of X-ray obscuring column. For BLR nondetections we plot
the ratio of an upper limit for a FWHM 3000\,km/s BLR to the total observed
(=narrow) Brackett~$\alpha$ flux. 
As qualitatively indicated, Seyfert 1 galaxies
would populate the upper left of this diagram and even larger ratios
of broad and narrow flux (e.g., Stirpe \cite{stirpe90}, median 20 in that
sample). The curved dotted line indicates the location of an object having 
an intrinsic broad to narrow line ratio of 20, with the broad component 
increasingly obscured by
dust with a Galactic ratio of infrared to X-ray obscuration. Changing adopted
dust properties would move this line horizontally, changing the adopted 
intrinsic ratio would move the line vertically.
}
\label{fig:deteclimits}
\end{figure*}

Another complex case is NGC\,5506 where a broad component to Paschen
$\beta$ has been reported (Blanco et al. \cite{blanco90}; 
Rix et al. \cite{rix90}). Goodrich et al. (\cite{goodrich94})
and Veilleux et al. (\cite{veilleux97}) confirm the larger width
of this infrared line compared to optical lines but, on the basis of
a similar [Fe\,II] profile, ascribe this
effect to a moderately broad but obscured component of the NLR.
Our profile of Brackett~$\alpha$ is relatively broad but not obviously
the sum of a narrow and a broad line. Comparison to [Si\,IX] does not
help in this case because of its weak and uncertain detection
in NGC\,5506, making the measured FWHM highly uncertain.  
We hence compared the Brackett~$\alpha$ profile with ISO spectroscopy 
of the [O\,IV] 26$\mu$m forbidden line in this
object (Sturm et al. \cite{sturm02}) which is able to 
penetrate large obscuring columns. We fixed the NLR FWHM for the ISAAC spectrum
to the value of 460\,km/s which is based on the FWHM measured from a 
gaussian fit to [O\,IV], corrected for the resolution difference of 
ISO-SWS and ISAAC.
Adopting this NLR line width we obtain a residual BLR component with
FWHM similar to the one of Blanco et al.
(\cite{blanco90}) but significantly lower than the value
reported by Rix et al. (\cite{rix90}), most likely reflecting line profile 
decomposition uncertainties of the various datasets. We also checked for
differences between the fine-structure line FWHM (Sturm et al. 
\cite{sturm02}) and the optical FWHM of Veilleux 
(\cite{veilleux91a}; \cite{veilleux91b}) that would be 
expected if extinction within the NLR
dominates the profiles. No significant differences were found when
convolving the optical data to the lower resolution and consistently
measuring the FWHM by gaussfits. On the basis of
an O\,I fluorescence line and of J band [Fe\,II] transitions, Nagar et al. 
(\cite{nagar02}) argue that NGC\,5506 is an obscured Narrow Line Seyfert 1,
consistent with our decomposition which ascribes most of Br$\alpha$ to
a fairly narrow BLR.

The most uncertain case is NGC\,2110 with its moderately wide 
Brackett~$\alpha$ line.
The nondetection of [Si\,IX] and the absence of ISO spectroscopy makes
a direct comparison to the forbidden lines impossible. 
Evidence on possible broad components to Paschen $\beta$ is mixed
(Rix et al. \cite{rix90}; Veilleux et al. \cite{veilleux97}).
We have listed
this case as a narrow line, on the basis of the similarity of the 
Brackett~$\alpha$ line width with that of [Fe\,II] (Veilleux et al. 
\cite{veilleux97}). We will include this object as uncertain when discussing
the sample statistics below.
    
Figure~\ref{fig:deteclimits} summarizes the detections and limits on broad
components of Brackett~$\alpha$ in our sample objects. Relatively few
broad line regions are found -- only 3 to 4 out of 13 (considering the 
uncertain
case of NGC\,2110). For the nondetections, limits for broad Br$\alpha$ 
components are of the order 1 to 2 times the narrow Br$\alpha$ flux. 
Assuming intrinsic presence of a BLR, as confirmed by
spectropolarimetry for most of the objects, and an
intrinsic broad to narrow line ratio 20 equal to the median of the 
Stirpe (\cite{stirpe90}) $M_{\rm V}<-22$ Seyfert 1s, 
these limits imply an obscuration of broad Brackett~$\alpha$ of $\approx$3\,mag.
For a Galactic extinction curve, this
corresponds to an equivalent visual extinction of more than 50 magnitudes.
For $A_{4.05}/A_{\rm V}$=0.035 (Draine \cite{draine89}) the equivalent visual 
extinction would be 86mag, for $A_{4.05}/A_{\rm V}$=0.051 (Lutz \cite{lutz99})
59 mag. Assuming a canonical conversion factor from visual extinction to X-ray
obscuring column of $N_{\rm H}/A_{\rm V}=1.79\times 10^{21}$\,cm$^{-2}$ 
(e.g. Predehl
\& Schmitt \cite{predehl95}), broad components to Brackett~$\alpha$ should
remain visible up to X-ray obscuring columns of about 10$^{23}$\,cm$^{-2}$.
The fact that all our BLR detections are below this limit is thus consistent
with a Galactic value of the ratio of 4$\mu$m obscuration to X-ray column. 
   
At this point, a comparison to the factors determining the detectability of 
Seyfert 2 BLRs in polarized light is in place.
The dominant factor for BLR detectability in polarized light 
is AGN luminosity, probably both through the corresponding variation of
host galaxy dilution and through a varying size of the scattering region 
(e.g. Alexander \cite{alexander01}; Lumsden et al.~\cite{lumsden01};
Lumsden \& Alexander \cite{lumsden01b}). 
The BLR detectability in polarized light
one modestly depends on the X-ray column (Lumsden et al.~\cite{lumsden01}), 
there are BLR detections up to the highest X-ray columns. 
Since most of our objects do have detections of the BLR in polarized 
light (Table~\ref{tab:fluxes}), and were selected to be bright in 
[O\,III], we believe that the effect of AGN luminosity and host dilution
on the BLR detectability in our spectroscopy is less important than
the effect of obscuration.

\section{Discussion}
\label{sect:discussion}
Our observations have resulted in the detection of broad line regions in
about one quarter of the Seyfert 2 sample studied. This fraction is 
similar to the one reported at shorter near-infrared wavelengths by 
Veilleux et al. (\cite{veilleux97}). Small numbers and the 
difficulties discussed above to ascribe moderately broad profiles 
for some of the objects to a BLR or the NLR limit a detailed
intercomparison of these fractions. It is clear, however, that
the detection rate at the wavelength of 4$\mu$m which samples
columns up to 10$^{23}$\,cm$^{-2}$ (for a Galactic extinction curve) 
is not much higher than in the shorter wavelength studies which penetrate 
3 to 5 times smaller columns. This cannot reflect intrinsic absence of BLRs
since they are detected through spectropolarimetry in the majority of the
objects with 4$\mu$m BLR nondetections. The majority of Seyfert 2s thus still 
has considerable BLR obscuration at 4$\mu$m. 
This finding is still consistent
with most popular models of dust around AGN, though being close to 
constraining the ones
that predict the most modest obscurations (e.g., Granato et al.
\cite{granato97}). It is already exceeding the modest
obscurations ($A_{\rm V}$ up to a few tens) predicted for most polar angles 
($\theta < 90\deg$) in the disk wind model of K\"onigl \& Kartje 
(\cite{koenigl94}).

The high 4$\mu$m obscuration and the consistency with a Galactic 
extinction curve raise an apparent conflict with the abnormally low
ratios of {\em visual} reddening $E_{\rm B-V}$ and X-ray column in luminous AGN,
as summarized by Maiolino et al. (\cite{maiolino01a}). Comparing reddening
$E_{\rm B-V}$ towards the BLR as measured from optical spectra with X-ray 
$N_{\rm H}$,
they find this ratio to be 3 to 100 times lower than Galactic for most of 
the objects in their sample. Based on other evidence, they suggest the ratio 
of visual
extinction $A_{\rm V}$ and $N_{\rm H}$ to be lower than Galactic as well, for several
classes of AGN. 
Simply lowering the dust content in the absorber and releasing the metals 
into the gas does not solve this discrepancy of about an order of 
magnitude. 
In addition to other effects discussed by Maiolino et al. (\cite{maiolino01b}), 
a low dust-to-gas ratio should result
in the detection of broad Brackett~$\alpha$ lines at X-ray columns 
approaching 10$^{24}$\,cm$^{-2}$, but none are observed in our sample.  

Geometry effects may play a role since the samples are different. Maiolino
et al. (\cite{maiolino01a}) mainly observed Seyfert~1s, Quasars and
intermediate Seyferts, since optical detection of a BLR was pre-requisite
for their analysis. Here, we analyze intermediate and type 2 objects,
that is ones viewed more `edge-on' in the picture of a central torus.
If, as not implausible, dust were preferentially modified or destroyed
along the opening of the torus, then different lines of sight probe 
different dust properties. 
The modified dust may be found detached in the opening of the torus, 
or associated with the walls.
Similarly, the dusty wind model of K\"onigl \& Kartje (\cite{koenigl94})
provides a geometry where dust-free gas is surrounded by dusty
gas but, as noted above, predicts relatively low obscuration.
Another geometric option is that `normal' dust on large scales (100\,pc or more)
contributes to the obscuration, as well as `modified' circumnuclear 
dust, the two covering different directions. This geometry is similar to the 
scenario proposed by Maiolino \& Rieke (\cite{maiolino95}) for
the RSA Seyferts, but with the difficulty that they invoke 
the likely more `normal' large scale dust to create the
obscuration of intermediate Seyferts. This conflicts with observations 
that intermediate Seyferts do show 
anomalous relations of optical and X-ray columns (Maiolino et al. 
\cite{maiolino01a}). This is not consistent with `normal' galactic dust, 
unless one
postulates that there is a spatial separation between X-ray absorber
(dust-free and close to the nucleus) and normal dust optical absorption
on large scale (Weingartner \& Murray~\cite{weingartner02}). 

\begin{figure}
\center{\resizebox{0.9\columnwidth}{!}{\includegraphics{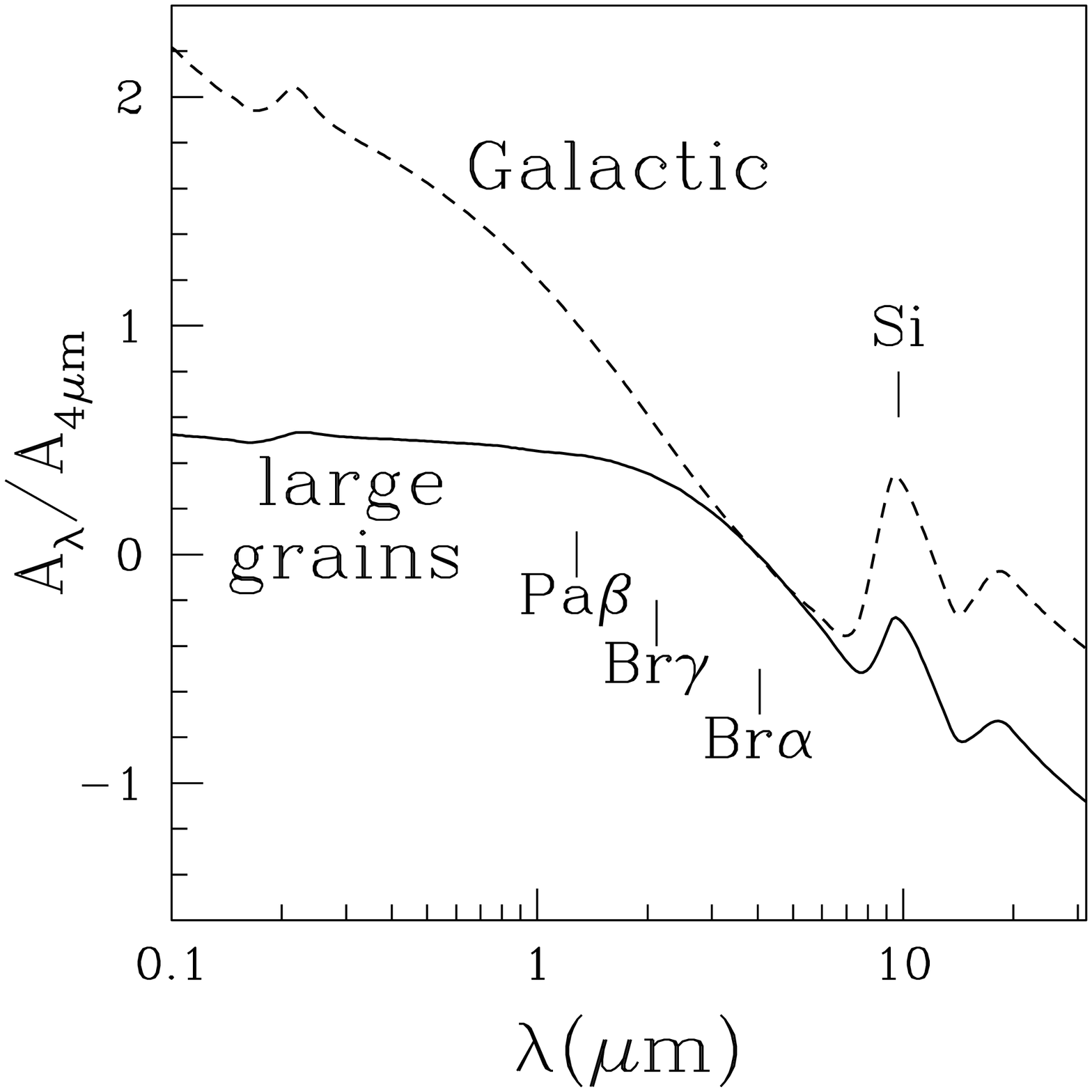}}}
\center{\resizebox{0.9\columnwidth}{!}{\includegraphics{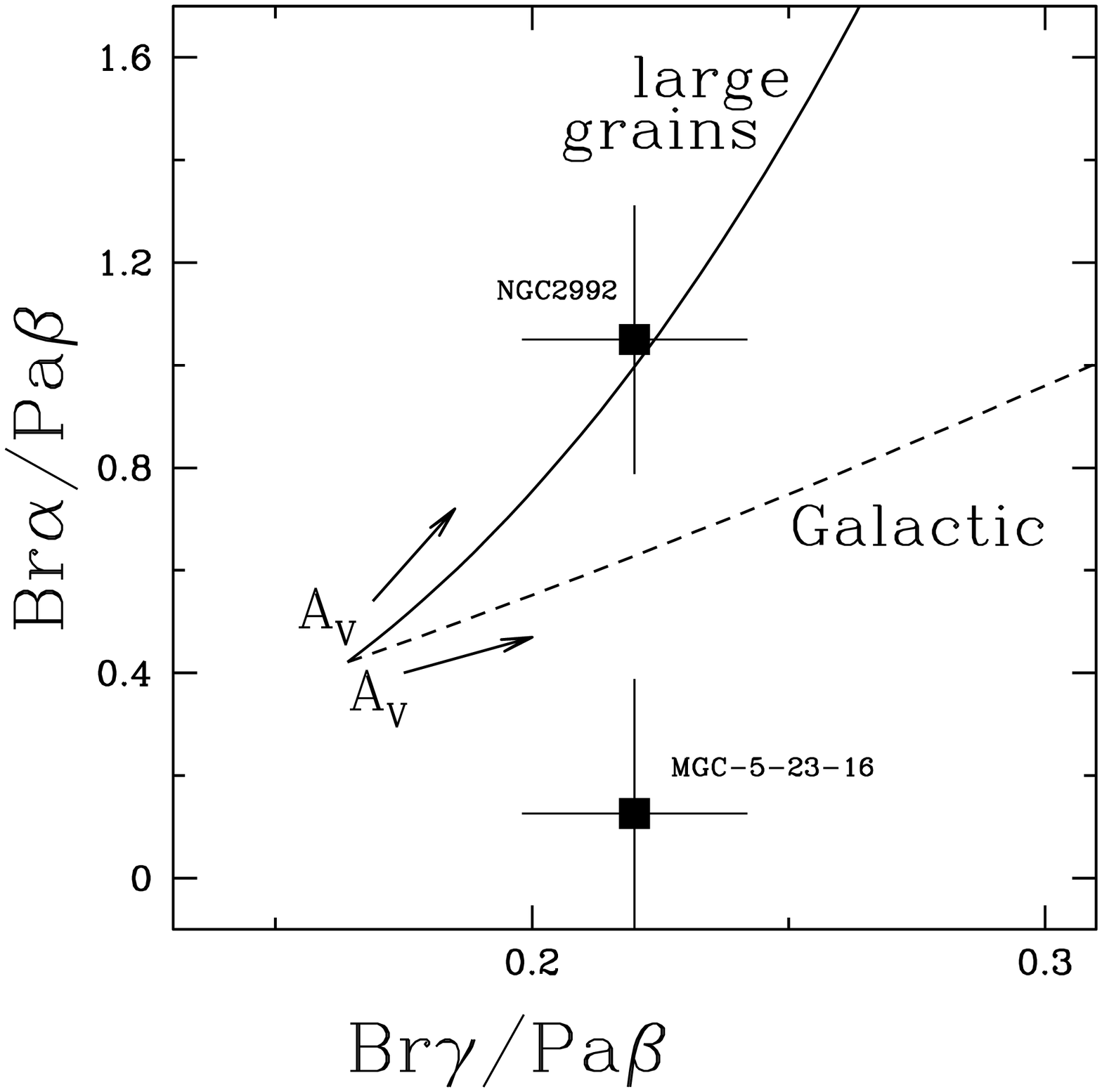}}}
\caption{Top: Changing the dust size spectrum from a standard `Galactic' one
to one dominated by large grains may flatten the extinction curve in the
optical and induce a knee in the near-infrared (see text for details).
Bottom: Expected near-infrared broad line ratios for increasing obscuration
and the two cases of standard Galactic grains and large grains. 
The adopted intrinsic ratios are for case B and T$_{\rm e}$=20000\,K, 
n$_{\rm e}$=10$^9$\,cm$^{-3}$.
The two objects from our Br$\alpha$ sample for
which {\em non-simultaneous} Pa$\beta$ and Br$\gamma$ data are available 
are indicated. Future
quasi-simultaneous and accurately calibrated spectroscopy is needed to 
actually execute this test.}
\label{fig:tworatios}
\end{figure}

The seemingly discrepant optical and  4$\mu$m results are, however, also
consistent with the preferred interpretation of Maiolino et al.
(\cite{maiolino01b}) for the results of Maiolino et al. (\cite{maiolino01a}).
If dust in the dense circumnuclear regions of AGN is dominated by
large grains, as proposed by Laor \& Draine (\cite{laor93}) and by Maiolino
et al. (\cite{maiolino01b}), the net effect on the extinction curve will be 
mainly a reduction at short wavelengths, but much less change in the 
infrared. This is illustrated in the top panel of Fig.~\ref{fig:tworatios}, 
where the Galactic `standard' extinction curve is compared with an extinction
curve due to a grain distribution biased in favor of large sizes
\footnote{More specifically, the distribution of grain sizes has been
modelled with $n\propto a^{-\beta}$, $\beta = 2.5$,
$a_{\rm min}=0.005{\rm \mu m}$, $a_{\rm max}=1{\rm \mu m}$, among those 
suggested by Laor
\& Draine (1993) and by Maiolino et al. (\cite{maiolino01b}),
at variance with the "standard" distribution for the Galactic dust,
where $\beta = 3.5$,
$a_{\rm min}=0.005{\rm \mu m}$, $a_{\rm max}=0.25{\rm \mu m}$. See also 
Maiolino et al. (\cite{maiolino01b}).}.
This effect would explain both the mismatch between optical extinction and
gaseous column measured in the X-rays (discussed in Maiolino et al. 
\cite{maiolino01a})
and the agreement between IR extinction and X-ray absorption
found in this paper.

The `large grains' curve has a turnover at about 2--3$\mu$m (at
variance with the `Galactic' curve) which is nicely probed by the three
infrared hydrogen lines Pa$\beta$, Br$\gamma$ and Br$\alpha$. In case of
(screen) absorption, the deviation of the line ratios
from the intrinsic case B depends significantly
on the extinction curve, this is shown in the lower panel of
Fig.~\ref{fig:tworatios}. For two
of the sources showing broad Br$\alpha$, namely NGC2992 and
MCG-5-23-16, there are measurements in the literature for the broad
components of Pa$\beta$ and Br$\gamma$ (Gilli et al. \cite{gilli00},
Veilleux et al. \cite{veilleux97}), which allow to locate these
objects in the lower panel of Fig.~\ref{fig:tworatios}. The location of 
NGC\,2992 is inconsistent with the
Galactic extinction curve, but fully consistent with the `large grains'
scenario. For MCG-5-23-16 the location is not even consistent with the
Galactic extinction, because of the very low Br$\alpha$/Pa$\beta$ ratio.
This might be due to variability of the BLR, since Br$\alpha$ (this
paper) and Pa$\beta$ (Veilleux et al. \cite{veilleux97}) were not measured 
simultaneously but with a time lag of 9 years. This point of view is
strengthened by the data of Blanco et al. (\cite{blanco90})
who, another 3 years earlier, measured a 4 times lower broad Pa$\beta$ flux.  
The caveat of variability might also apply to
NGC\,2992, though here the measurements were much closer in time (2 years) and
the data from Gilli et al. (2000) have higher accuracy.
Additional, simultaneous measurements of the broad components of the infrared 
hydrogen
lines are required to unambiguously test the `large grains' scenario.

We note that a population of large grains in the dense circumnuclear
regions of AGN is not an ad-hoc requirement, but in line with the flattened
extinction curves in Galactic dense molecular clouds (eg.
Cardelli et al. 1989) which are ascribed to grain coagulation. However,
grain growth may yield the formation of complex fluffy
aggregrates (e.g., Dominik \& Tielens \cite{dominik97}) rather than the 
simply larger but spherical grains, which are assumed in the 
simplified model of Fig.~\ref{fig:tworatios}.
Finally, a dust distribution biased for large grains may also come from
the preferential destruction of small grains by sublimation or sputtering
in the circumnuclear region of AGN. Depending on the mechanism producing a 
dust distribution that is biased toward large grains, the {\em absolute}
extinction $A_{4\mu m}/N_{\rm H}$ may differ. In case of destruction of small
grains it will be below but similar to the Galactic value, while for 
coagulation at a fixed dust mass it will rise above the Galactic value.
Fig.~\ref{fig:deteclimits} is consistent with modest deviations in either
direction.

The current data do not allow to discriminate between the `geometrical' and
the `large grains' scenario. Both are plausible within
the unified scenario for AGN. A 
better understanding, at least for the intermediate type
Seyferts with moderate $A_{\rm V}$ could be gained through simultaneous observations
of the main optical to 4$\mu$m recombination lines, in order to directly
trace the flattening of the extinction curve expected in the large grain
scenario. Line ratios formed by
the broad components of Brackett~$\alpha$, Brackett~$\gamma$, and 
Paschen~$\beta$ (Fig.~\ref{fig:tworatios}) are better suited than a similar diagram
invoking the Balmer decrement, because of the reduced susceptibility to 
departures from case B (e.g. Netzer \cite{netzer90} and references
therein). While Fig.~\ref{fig:tworatios} proposes a direct
test for the presence of large grains, future quasi-simultaneous and 
accurately calibrated data will be needed for a conclusive result. 

\section{Conclusions}
\label{sect:conclusions}

We have presented new 4$\mu$m spectroscopy of a sample of 12 Seyfert 2 galaxies
that are well-studied in the X-ray, and combine these data with previous 
spectroscopy of NGC\,1068. The observations are designed
to probe for the presence of optically obscured Broad Line Regions. 
The main results are\\
(i) Broad components  to Brackett~$\alpha$ are detected in 3 to 4 of these 13
galaxies.\\
(ii) The detections and limits are consistent with a Galactic ratio of
infrared and X-ray obscuring columns. This result can be reconciled with the
low ratios of optical to X-ray obscuring columns observed for several AGN
if either the obscuring dust consists of large grains leading to a 
modified extinction
curve, or if variation in dustiness or dust properties exists  between 
directions probing right through the putative torus and directions closer 
to its opening.\\
(iii) A survey of the coronal [Si\,IX] 3.94$\mu$m line shows considerable
variation in its ratio to Brackett~$\alpha$.\\
(iv) Two coronal lines of [Ca\,VII] and [Ca\,V] are detected for the first 
time in an extragalactic object, the Circinus galaxy.

\begin{acknowledgements}
We thank the Paranal staff for excellent support, Bill Vacca for 
discussions, and the referee for helpful comments. This work is supported 
by GIF grant I-551-186.07/97.
\end{acknowledgements}

\end{document}